\documentclass[%
reprint,
showpacs,
 amsmath,amssymb,
 aps,
pra,
]{revtex4-1}

\usepackage{graphicx}
\usepackage{dcolumn}
\usepackage{bm}

\usepackage[caption=false,font=footnotesize]{subfig}
\begin{document}

\title{Space-Time Cross-Mapping\\and Application to Wave Scattering}

\author{Mohamed A. Salem}
 \email{mohamed.salem@polymtl.ca}
\author{Christophe Caloz}%
 \email{christophe.caloz@polymtl.ca}
\altaffiliation[Also with ]{King Abdulaziz University, Jeddah, Saudi Arabia.}%
 \affiliation{ Poly-Grames Research Center, Polytechnique Montr\'{e}al, Montr\'{e}al, Qu\'{e}bec, H3T 1J4 Canada}
%


\date{\today}

\begin{abstract}
Causality creates an asymmetry between space and time, even though the wave equation treats them on equal footing. In this work, we leverage this asymmetry to construct a cross-mapping between space and time. This cross-mapping is applied to simplify scattering in space-varying media, by eliminating the infinite reflections between interfaces. The method is shown to transform the implicit transfer matrix method into an explicit method for the analysis of electromagnetic field scattering by a stratified medium.
\end{abstract}

\pacs{41.20.Jb, 42.25.Gy, 03.50.De}
\maketitle


\section{Introduction}

Time-varying media alter the angular frequency spectrum of waves propagating across them. Such transformations are rarely discussed in electrodynamics textbooks, except for the Doppler effect, which is essentially a change in frequency due a moving source or scattering medium as seen by a stationary observer. However, time-varying media systems not involving any motion, such as magnetoplasma and magneto-elastic media, may be more practical in electromagnetic applications~\cite{Auld:68,Kong:75,Biancalana:07,Kalluri:10}. Related effects include frequency conversion, time reversal, and angular spectrum engineering possibilities, with a wide range of potential applications in electromagnetics and optics. 

In this work, we present a space-time cross-mapping approach where scattering in space-varying media is treated as scattering in corresponding time-varying media. This approach may simplify the analysis of space-varying scattering problems by leveraging the breaking, due to causality, of the inherent space-time symmetry in the wave equation. Indeed, while the wave equation admits solutions that correspond to waves propagating forward and backward in space and in time, however, causality prevents temporal backward-propagating waves. In the case of discontinuous time-varying media, such abruptly switched media, this breaking of space-time symmetry prohibits multiple reflection between the temporal interfaces. Treatment of wave scattering in such media is inherently simpler than in corresponding space-varying media with multiple reflections between the interfaces. However, certain challenges also arise in the cross-mapped configuration, such as the presence of boundaries outside the line-of-sight (LoS) of the source and observer. This challenge arises precisely due to space-time asymmetry, where boundaries outside the LoS in a time-varying configuration are not permitted to reflect waves backward in time. A second closely related challenge is the proper cross-mapping of sources from spatial into temporal configurations.

This work is organized as follows: In Section~\ref{sec:2}, the principles of wave scattering at temporal discontinuities, which are the simplest form of time-varying media, are detailed. The relevant continuity conditions, conservation laws and an ongoing controversy pertaining to the electromagnetic wave momentum are discussed and canonical scattering at simple temporal discontinuities are analyzed. In Section~\ref{sec:3}, the space-time cross-mapping method is formulated. First, the appropriate mapping between the incident and scattered waves in temporal and spatial configurations, and the relevant mapping of media properties are established. Next, the treatment of corresponding boundary and continuity conditions is presented, followed by the mapping of sources. In Section~\ref{sec:4}, applying the cross-mapping is shown to simplify the transfer matrix method (TMM) by transforming it from an implicit into an explicit method. Conclusions are given in Section~\ref{sec:5}.

\section{Wave Scattering at Temporal Discontinuities}
\label{sec:2}

Wave scattering at spatial discontinuities has been thoroughly covered in the literature. Examples of canonical scattering at simple spatial discontinuities include the problems of two half-spaces, slabs, and stratified media~\cite{Kong:75}. Understanding canonical scattering at simplified discontinuities provides a basis for understanding more involved scattering phenomena. By analogy, understanding wave scattering at temporal discontinuities is the basis for understanding more involved temporal scattering phenomena and may be fundamental for complete understanding of electrodynamic phenomena in general. 

In all cases, understanding wave scattering at an interface requires establishing continuity conditions through the interface. In what follows, we show the difference between the continuity conditions at spatial and temporal interfaces, and apply these conditions to scattering by simple temporal discontinuities.

\subsection{Continuity Conditions and Conservation Laws}

In the scattering of electromagnetic fields at a spatial interface, the conventional continuity conditions imply the continuity of the tangential electric and magnetic fields at the interface. Even though the mathematical formulation of these conditions is not rigorous in the case of discontinuities~\cite{Schelkunoff:72,Idemen:11}, it yields satisfactory results away from the discontinuity in most practical configurations. However, a more rigorous interpretation of the continuity conditions in terms of actual physical principles is still preferable, especially when investigating unconventional configurations. The physical principles involved are conservation laws, such as the conservation of energy and momentum across the discontinuity.

Scattering at passive and lossless spatial discontinuities implies the conservation of energy flow, whence all incident power is transformed into scattered power. In the case of harmonic waves, the energy conservation condition reads%
\[
\langle \mathbf{S}^{\mathrm{i}}\rangle = \langle \mathbf{S}^{\mathrm{r}}\rangle + \langle \mathbf{S}^{\mathrm{t}}\rangle,
\]
where $\langle\mathbf{S}\rangle$ is the time-averaged power density vector, and the superscripts $\mathrm{i}$, $\mathrm{r}$ and $\mathrm{t}$ respectively refer to the incident, reflected and transmitted waves. In addition to the conservation of energy, in linear media the \textit{rhythm} of the wave, i.e. its temporal dependence, is also conserved to satisfy the phase matching at the discontinuity. This phase matching condition is what dictates the change in the wave vector of the scattered wave. Let a harmonic wave propagating in the $z$-direction be given by $\exp(i [\pm k_\mathrm{j} z \pm \omega_\mathrm{j} t])$, where $\mathrm{j} \in \{\mathrm{i}, \mathrm{r}, \mathrm{t}\}$. Assuming an incident wave propagating in the positive $z$-direction, $\exp(i [k_\mathrm{i} z - \omega_\mathrm{i} t])$, the phase matching condition at a discontinuity located at $z=0$ reads%
\begin{equation}
\label{eq:spatial_disc}
e^{-i \omega_\mathrm{i} t} + \gamma e^{- i \omega_\mathrm{r} t} = \tau e^{- i \omega_\mathrm{t} t},\quad \forall t,
\end{equation}
where $\gamma$ and $\tau$ are the coefficients of the scattered waves in the incidence medium and scattering medium, respectively. Condition~(\ref{eq:spatial_disc}) thus implies the conservation of the angular frequency%
\begin{equation}
\label{eq:ph_s}
\omega_\mathrm{i} = \omega_\mathrm{r} = \omega_\mathrm{t}.
\end{equation}
For simplicity, the signs of $\omega_\mathrm{j}$ are chosen to be the same. Consequently, the wavevectors of the reflected and transmitted waves have to change to satisfy the wave dispersion relation, $\omega_\mathrm{j}^2 = k_\mathrm{j}^2 c^2$, where $c$ is the speed of light in free-space. This implies that%
\begin{equation}
\label{eq:ph_sk}
\frac{k_\mathrm{i}}{ n_\mathrm{i}} = -\frac{k_\mathrm{r}}{n_\mathrm{r}} = \frac{k_\mathrm{t}}{n_\mathrm{t}},
\end{equation}
where $n_\mathrm{j}$ is the refractive index in the corresponding medium.

Alternatively, at a temporal discontinuity, the phase matching condition results in conservation of the \textit{pattern} of the wave, i.e. the wavevector in the case of harmonic waves. Assuming an incident wave propagating in the positive $z$-direction, the phase matching condition at a temporal discontinuity occurring at $t=0$ reads%
\begin{equation}
\label{eq:temporal_disc}
e^{i k_\mathrm{i} z} = \xi e^{ i k_\mathrm{f} z} + \overline{\xi} e^{ i k_\mathrm{b} z},\quad\forall z,
\end{equation}
where $\xi$ and $\overline{\xi}$ are the coefficients of the scattered waves in the medium after the discontinuity. Here, it is essential to notice that, due to causality, no wave is permitted to reflect in time and propagate into the incidence medium, in contrast to the case of Eq.~(\ref{eq:spatial_disc}). The condition~(\ref{eq:temporal_disc}) thus implies the conservation of the wavevector%
\begin{equation}
\label{eq:ph_t}
k_\mathrm{i} = k_\mathrm{f} = k_\mathrm{b}.
\end{equation}
Similarly, for simplicity, the signs of $k_\mathrm{j}$ are chosen to be the same. Consequently, the angular frequencies of the scattered waves have to change to satisfy the wave dispersion relation. This implies that%
\begin{equation}
\label{eq:ph_tw}
\omega_\mathrm{i} n_\mathrm{i} = \omega_\mathrm{f} n_\mathrm{f} = -\omega_\mathrm{b} n_\mathrm{b}.
\end{equation}
It is also worth noting that negative sign in front of $\omega_\mathrm{b}$ means that the wave is propagating in the negative $z$-direction, since the wavenumber has not changed according to Eq.~(\ref{eq:ph_t}). Accordingly, at a temporal discontinuity, an incident wave splits into a forward- and a backward-propagating waves while incurring shift in the angular frequency.

The conservation of $k$ suggests that corresponding physical conservation law invoked at a temporal discontinuity is the conservation of momentum flow of the wave. This implies that the momentum flow of the incident wave must be equal to the momentum flow of the scattered wave at the discontinuity, which splits the scattered wave into a forward- and backward-propagating waves. The continuity condition for a space-harmonic wave may thus be expressed as%
\[
\langle\mathbf{G}^{\mathrm{i}}\rangle = \langle\mathbf{G}^{\mathrm{b}}\rangle + \langle\mathbf{G}^{\mathrm{f}}\rangle,
\]%
where $\langle\mathbf{G}\rangle$ is the space-averaged momentum density.

The definition of the momentum density of electromagnetic waves in a dielectric medium is a controversial issue. Two expressions exist with claimed theoretical and experimental support and still no conclusive resolution. The two expressions are derived from the photon momentum in a dielectric medium with refractive index $n$. In Ref.~\cite{Minkowski:10}, Minkowski derived the (canonical) photon momentum as $p_{\mathrm{M}} = n \hbar \omega/c$, where $\hbar$ is the reduced Planck constant, yielding the electromagnetic momentum density $\mathbf{G}_{\mathrm{M}} = \mathbf{D} \times \mathbf{B}$, where $\mathbf{D}$ and $\mathbf{B}$ are the displacement field and the magnetic flux, respectively. On the other hand, in Ref.~\cite{Abraham:09}, Abraham derived the (kinetic) photon momentum as $p_{\mathrm{A}} = \hbar \omega / (n c)$, yielding the electromagnetic momentum density $\mathbf{G}_{\mathrm{A}} = \frac{1}{c^2} \mathbf{E} \times \mathbf{H}$, where $\mathbf{E}$ and $\mathbf{H}$ are the electric and magnetic fields, respectively. This difference in electromagnetic momentum expressions resulted in what is known as the Abraham-Minkowski controversy~\cite{Pfeifer:07,Pfeifer:09}. 

Since the Abraham-Minkowski controversy is still not resolved in a definite fashion, the proposed cross-mapping method is presented for scalar waves. This scalar representation avoids ambiguity in the correct form of the continuity condition at temporal discontinuities, since the expressions for the energy and momentum densities are unambiguous for scalar waves. This representation however does not limit the applicability of the proposed method, since what is actually modeled is a spatial scattering problem, not a temporal scattering one, and the correct continuity conditions are therefore well-defined. Hence, the results of the scalar method are directly applicable to the vector electromagnetic case with proper translation as detailed in Section~\ref{sec:3}.

\subsection{Simple Discontinuities}
\label{sec:2simple}

To understand the space-time cross-mapping method, it is necessary to first study scattering at simple temporal discontinuities, such as single interfaces, temporal slabs, and stratified temporal media. In what follows, we continue with the example of plane-wave incidence on a temporal discontinuity in the previous section. The phase matching condition implied that $k$ is conserved and $\omega$ is changed in the scattering medium. To determine the scattered wave coefficients, the continuity of the time derivative of the wave across the discontinuity is invoked, which translates to the continuity of the momentum density in the vectorial case. The continuity of time derivative of the wave reads%
\begin{equation}
\label{eq:temporal_flow}
-i \omega_\mathrm{i} e^{i k_\mathrm{i} z} = -i \omega_\mathrm{f} \xi_\mathrm{f} e^{i k_\mathrm{i} z} - i \omega_\mathrm{b} \xi_\mathrm{b} e^{i k_\mathrm{i} z},\quad\forall z.
\end{equation}
Conditions~(\ref{eq:temporal_disc}) and~(\ref{eq:temporal_flow}) with the relations~(\ref{eq:ph_t}) and~(\ref{eq:ph_tw}) determine the scattered wave coefficients as%
\begin{align}
\xi &= \frac{n_\mathrm{i} + n_\mathrm{f}}{2 n_\mathrm{i}},\label{eq:xi_f} \\
\overline{\xi} &= \frac{n_\mathrm{i} - n_\mathrm{f}}{2 n_\mathrm{i}}. \label{eq:xi_b}
\end{align}

\begin{figure}[ht]
\centering
\includegraphics[width=3.25in]{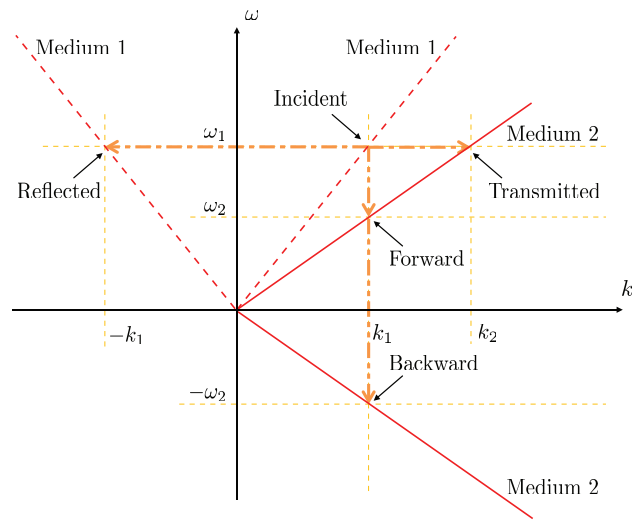}
\caption{Wave frequency shift at an interface between two media. At a spatial interface, the wave number of the incident wave is shifted (dash-dotted arrow) in the reflected and transmitted waves, while at a temporal interface, the angular frequency of the incident wave is shifted (dash-double-dotted arrow) in the forward- and backward-propagating waves.}
\label{fig:simple_interface}
\end{figure}

Figure~\ref{fig:simple_interface} depicts the frequency shift at simple spatial and temporal discontinuities. The slopes of the world lines (the straight lines passing through the origin) in the $\omega$--$k$ diagram designate the phase velocities in the two media. Conservation of $\omega$ at the spatial discontinuity, as indicated in Eq.~(\ref{eq:ph_s}), is designated by the horizontal dashed arrows. The $k$ values at the intersection points of the arrows with the world lines give the wavenumbers of the scattered waves, as indicated in Eq.~(\ref{eq:ph_sk}). Conservation of $k$ at the temporal discontinuity, as indicated in Eq.~(\ref{eq:ph_t}), is designated by the vertical arrows. The $\omega$ values at the intersection points of the arrows with the world lines give the angular frequencies of the scattered waves, as indicated in Eq.~(\ref{eq:ph_tw}).

\begin{figure}[ht]
\centering
\subfloat[]{\includegraphics[width=3.25in]{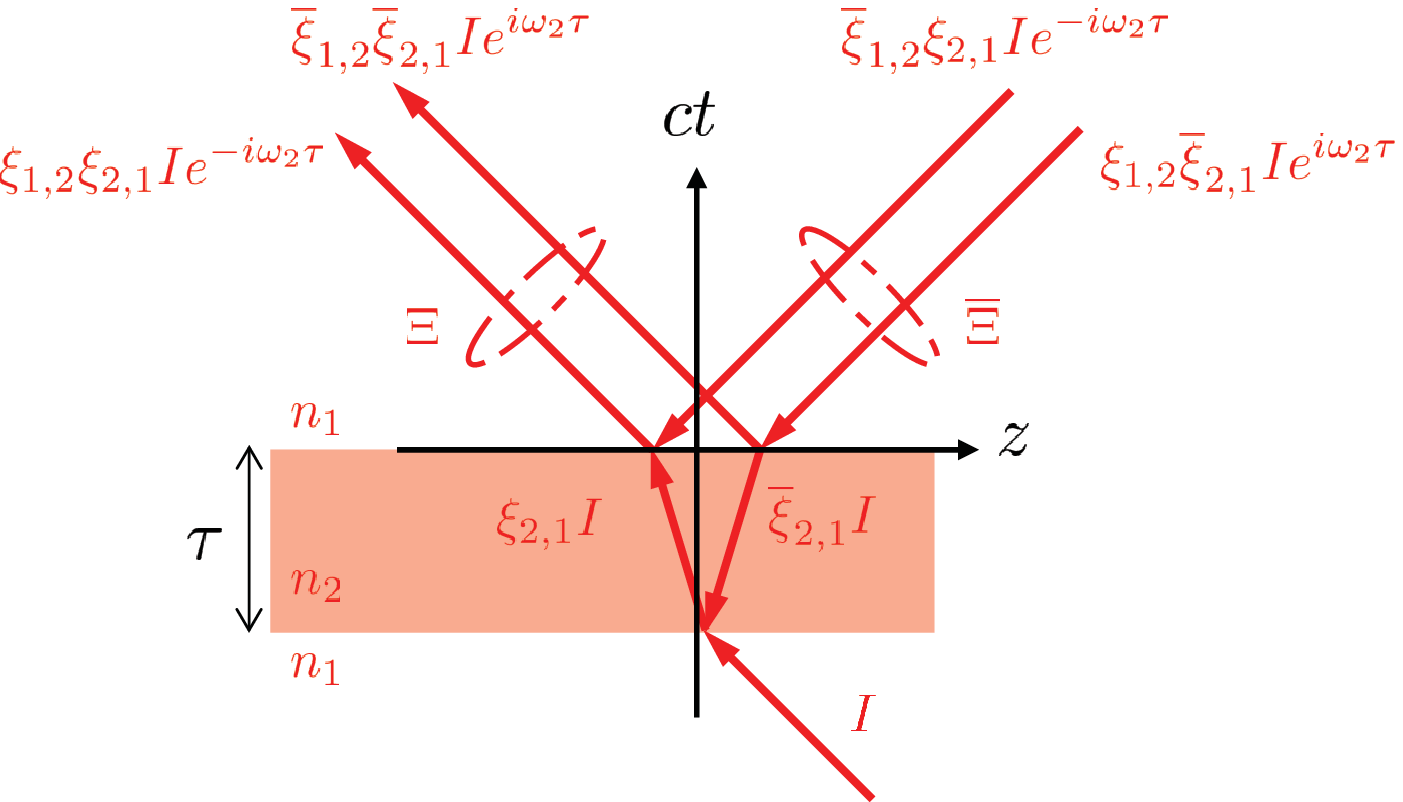}%
\label{fig:slaba}}
\\
\subfloat[]{\includegraphics[width=3.25in]{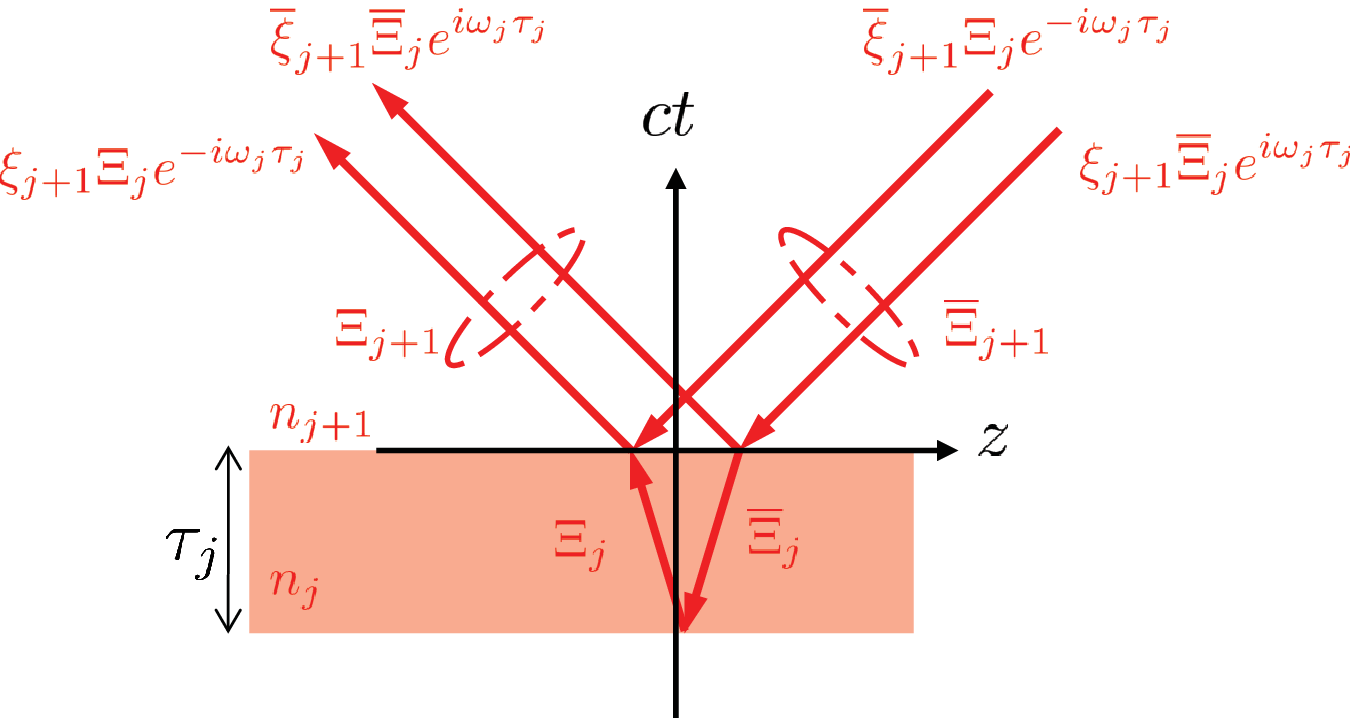}%
\label{fig:slabb}}
\caption{Wave scattering in temporal layered media. \protect\subref{fig:slaba} In a temporal slab, the composite scattering coefficients may be directly deduced from the local scattering coefficients at the two temporal interfaces of the slab. \protect\subref{fig:slabb} In a general multilayer temporal medium, the composite forward- and backward-propagating wave coefficients at a temporal interface may be directly deduced from the local scattering coefficients and the wave coefficients in the prior layer.}
\label{fig:slab}
\end{figure}

A natural extension to the temporal discontinuity is the temporal slab. A temporal slab is constructed by switching an unbounded medium and then switching this medium back to the initial medium after a time interval $\tau$ in between. Wave scattering through such slab follows as a direct extension of the discontinuity case. The waves scattered by the second slab interface are causal. Therefore, they are not allowed to reflect back in time and interact with the first interface. This is a manifestation of the asymmetry between space and time due to causality. Assuming a slab of refractive index $n_2$ and duration $\tau$, embedded in a background medium with refractive index $n_1$, the composite (global) coefficients of the forward- and backward-propagating waves after passing through the temporal slab are found by applying Eqs.~(\ref{eq:xi_f}) and~(\ref{eq:xi_b}) at each one of the two interfaces of the slab, yielding%
\begin{align}
\Xi &= e^{i \omega_1 \tau} \left[\xi_{2,1} \xi_{1,2} e^{-i \omega_2 \tau} + \overline{\xi}_{2,1} \overline{\xi}_{1,2} e^{i \omega_2 \tau} \right], \label{eq:xi_f_s}\\
\overline{\Xi} &= e^{-i \omega_1 \tau} \left[\xi_{2,1} \overline{\xi}_{1,2} e^{-i \omega_2 \tau} + \overline{\xi}_{2,1} \xi_{1,2} e^{i \omega_2 \tau} \right], \label{eq:xi_b_s}
\end{align}%
where the subscripts $\mathrm{p},\mathrm{q}$ designate the transition across a temporal discontinuity from a medium with $n_\mathrm{p}$ into a medium with $n_\mathrm{q}$. Figure~\ref{fig:slaba} shows a pictorial diagram of wave scattering through a temporal slab indicating local and composite scattered wave coefficients.

Figure~\ref{fig:rot} compares wave scattering through spatial and temporal slabs. In the case of spatial slabs, the composite reflection and transmission coefficients, $\Gamma$ and $T$, respectively, combine an infinite series of reflections and transmissions, while in the case of temporal slabs, the composite forward- and backward-propagation coefficients, $\Xi$ and $\overline{\Xi}$, respectively, are each a sum of only two waves. The difference between the two material slabs is due to causality, which prevents the waves from reflecting from temporal interfaces into the past. It is also worthwhile noting that the direction of the trajectories of harmonic waves do not pertain to energy transfer, but rather to the direction of phase evolution, as required by the phase matching condition~(\ref{eq:temporal_disc}).

Extending the temporal slab results to stratified (multilayer) temporal media and temporal photonic crystals is straightforward. The coefficients of forward- and backward-propagating waves for each additional temporal interface are recursively constructed from those at the previous one in the same fashion Eqs.~(\ref{eq:xi_f_s}) and~(\ref{eq:xi_b_s}) are constructed from Eqs.~(\ref{eq:xi_f}) and~(\ref{eq:xi_b}), as shown in Fig.~\ref{fig:slabb}. Thus, the forward- and backward-propagating wave coefficients in layer $j+1$ are constructed from the those in layer $j$ as%
\begin{align}
\Xi_{j+1} &= \xi_{j+1} \Xi_{j} e^{-i \omega_j \tau_j} + \overline{\xi}_{j+1} \overline{\Xi}_{j} e^{i \omega_j \tau_j} , \label{eq:xif_rec} \\
\overline{\Xi}_{j+1} &= \overline{\xi}_{j+1} \Xi_{j} e^{-i \omega_j \tau_j} + \xi_{j+1} \overline{\Xi}_{j} e^{i \omega_j \tau_j}. \label{eq:xib_rec}
\end{align}

\section{Cross-Mapping Formulation}
\label{sec:3}

As demonstrated in Section~\ref{sec:2}, wave scattering at temporal discontinuities is simpler than that at spatial discontinuities due to causality, which prevents reflection into the past. The proposed space-time cross-mapping method aims at leveraging this asymmetry between spatial and temporal discontinuities to simplify the computation of scattered waves in a space-varying environment. The proposed method thus cross-maps space and time coordinates, allowing for the treatment of scattering at spatial discontinuities as scattering at temporal discontinuities. That is, the spatial configuration is mapped into the corresponding temporal configuration, solved in this simpler configuration, and then mapped back to the initial configuration. This mapping aims at reducing the computational complexity, since no infinite reflections and transmissions are present in a temporal scattering configuration with a finite number of discontinuities. Moreover, resembling marching-on time schemes~\cite{Chew:01,Taflove:05}, albeit in space, the mapping may be employed in developing marching-on in space computational techniques.

Two particular issues require careful consideration in applying the method. The first is accounting for continuity and boundary conditions at discontinuities outside the source-observer LoS, and the second is accounting for the presence of sources and their emitted backward-propagating waves. These two aspects have no direct counterparts in temporal configurations due to causality. The boundaries outside the LoS in a temporal configuration have no effect on the scattered wave, since waves cannot scatter back into the past to reach the observer. Similarly, in the case of a source, the backward-propagating wave emitted by the source has no direct equivalent in a temporal configuration.

In the following sub-sections, first, space-time cross-mapping is formulated for source-free configurations without discontinuities outside the LoS. Next, the cross-mapping method is extended to account for the presence of discontinuities outside the LoS and the presence of sources, which completes the formulation of the method.

\subsection{Space-Time Cross-Mapping}
\label{sec:3a}

\begin{figure}[ht]
\centering
\subfloat[]{\includegraphics[width=2.25in]{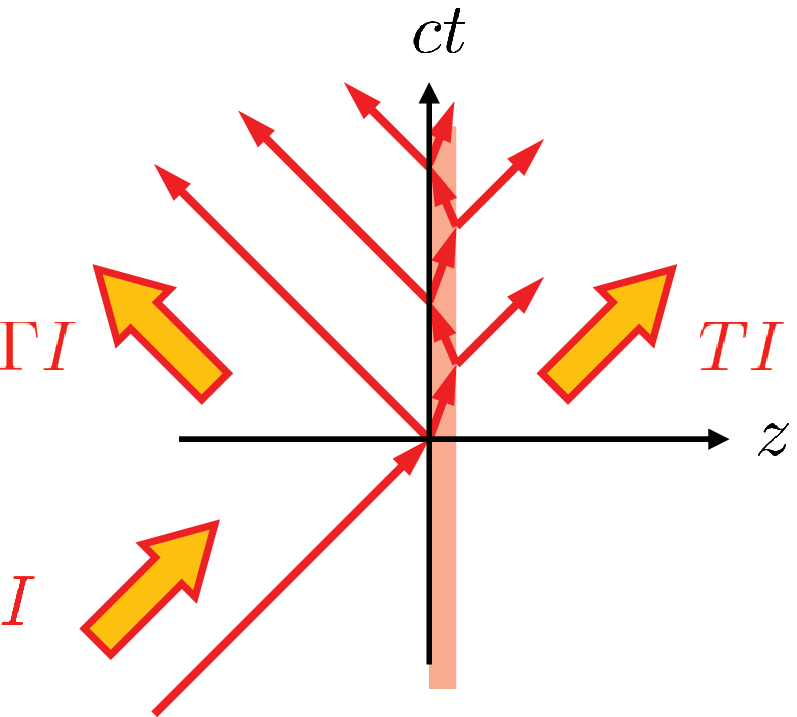}%
\label{fig:rota}}
\\
\subfloat[]{\includegraphics[width=2.25in]{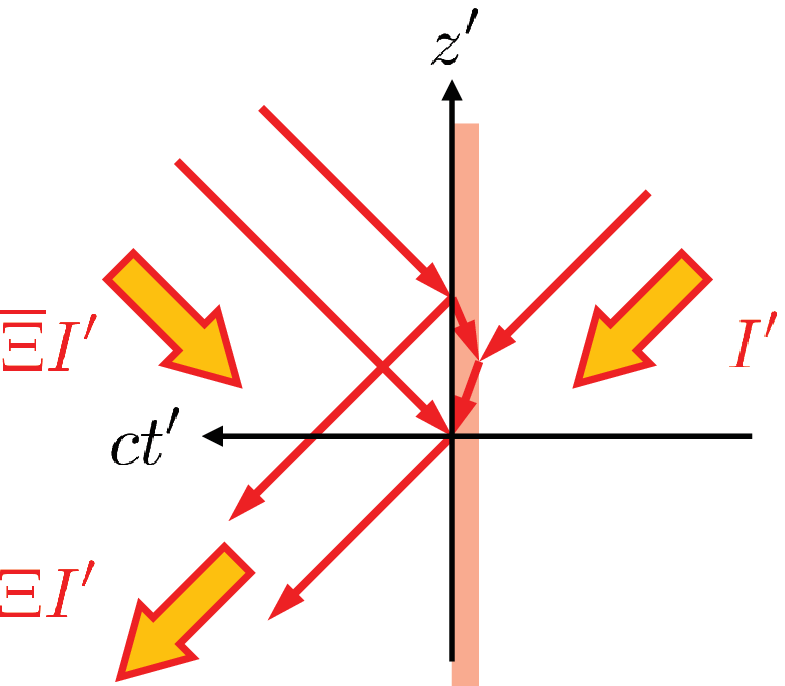}%
\label{fig:rotb}}
\caption{Dielectric slab cross-mapping between spatial configuration and temporal configuration. \protect\subref{fig:rota} Spatial configuration of the dielectric slab showing the incident, the infinite number of reflected and transmitted waves, and their composite representations, $I$, $\Gamma$, and $T$, respectively.\protect\subref{fig:rotb} Temporal configuration of the dielectric slab showing the incident, two forward- and two- backward-propagating waves, and their composite representations, $I'$, $\Xi$, and $\overline{\Xi}$, respectively. Note that in~\protect\subref{fig:rota}, the incident wave is propagating in the positive $z$-direction, while in~\protect\subref{fig:rotb}, the incident wave is propagating in the negative $z'$-direction. The arrows in both illustrations pertain to the evolution of the phase.}
\label{fig:rot}
\end{figure}

Consider the two cases of plane-wave scattering by a spatial material slab, shown in Fig.~\ref{fig:rota}, and by a temporal material slab, shown in Fig.~\ref{fig:rotb}. In the spatial configuration, the incident wave propagates in the positive $z$-direction, while, in the temporal configuration, the incident wave is considered to propagate in the negative $z$-direction. The geometry of the two scattering configurations are seen to be identical if one considers the composite wave coefficients rather than the individual trajectories, and inverts the direction of all the wave trajectories. This suggests that a mathematical mapping may by established between the two configurations.

The space-time cross-mapping procedure is thus carried out, with the aid of Fig.~\ref{fig:rot}, as follows: first, the spatial configuration coordinate axes are mapped as $z \rightarrow -ct'$ and $ct \rightarrow z'$. Second, the wave trajectories are reversed, resulting in mapping the transmitted wave in the spatial configuration into the incident wave in the cross-mapped temporal configuration. Third, the scattered waves in the cross-mapped temporal configuration are determined employing temporal treatment. Fourth, the substitution $n \rightarrow 1/n$ in applied to the final expressions of the composite wave coefficients. This substitution is necessary to account for the difference in frequency shift between spatial and temporal configurations, as shown in Eqs.~(\ref{eq:ph_sk}) and~(\ref{eq:ph_tw}). Finally, the obtained composite coefficients of the scattered waves are normalized with respect to composite forward-wave coefficient, $\Xi$. Table~\ref{tab:crossmap} provides a summary of cross-mapping for source-free configurations.

\begin{table}[b]
\caption{\label{tab:crossmap}%
Cross-mapping between spatial and temporal configurations}
\begin{ruledtabular}
\begin{tabular}{ccc}
Spatial & { } & Temporal \\
\hline
  $z$ & $\longleftrightarrow$ & $-ct'$ \\
  $ct$ & $\longleftrightarrow$ & $z'$ \\
  $n$ & $\longleftrightarrow$ & $1/n$ \\
  \hline
  $I$ & $\longleftrightarrow$ & $-I' \Xi$ \\
  $\Gamma I$ & $\longleftrightarrow$ & $-I' \overline{\Xi}$ \\
  $T I$ & $\longleftrightarrow$ & $-I'$ \\
  \hline
  $\Gamma$ & $\longleftrightarrow$ & $\overline{\Xi}/\Xi$ \\
  $T$ & $\longleftrightarrow$ & $1/\Xi$ \\
\end{tabular}
\end{ruledtabular}
\end{table}

To illustrate the cross-mapping procedure, consider the case of scattering through a dielectric slab of width $l$ with refractive index $n_2$ embedded in a background medium with refractive index $n_1$. Following Table~\ref{tab:crossmap}, and using Eqs.~(\ref{eq:xi_f_s}) and~(\ref{eq:xi_b_s}) lead to the following relation%
\begin{align*}
\Xi &= \frac{1 - \gamma^2 e^{2 i n_2 k l}}{\left[1 - \gamma^2 \right] e^{i n_2 k l}} = \frac{1}{T},\\
\overline{\Xi} &= \frac{\gamma \left[1 - e^{2 i n_2 k l} \right]}{\left[1 - \gamma^2 \right] e^{i n_2 k l}} = \frac{\Gamma}{T},
\end{align*}%
where $\gamma = (n_1 - n_2)/(n_1 + n_2)$. Now $\Gamma$ and $T$ may be directly computed without needing to sum up an infinite number of terms. The spatial scattering process has thus been simplified, by leveraging the asymmetry due to causality.

Note that space-time cross-mapping is not limited to harmonic waves and may be extended to pulsed waves in a straightforward fashion. A pulsed wave may be expanded in terms of its temporal spectral components through Fourier transform, assuming all media are linear, then the cross-mapping is carried out for each spectral component of the pulse. Finally, spectral composition through inverse Fourier transform of the spectral components of the scattered waves is carried out to determine the composite scattered waves.

As previously mentioned, cross-mapping is not restricted to scalar waves. It may directly be translated to vectorial electromagnetic configurations in the cases of longitudinally-invariant propagating modes and transverse electric-magnetic (TEM) fields. The cross-mapping in the vectorial case requires the substitution $n \rightarrow Z$ instead of $n \rightarrow 1/n$, where $Z$ is the impedance of the medium in the case of TEM fields, or the transverse impedance in the case of longitudinally invariant modes.

\subsection{Discontinuities outside the Line-of-Sight}

Discontinuities existing between the incident wave and the observer are naturally accounted for after applying the cross-mapping. However, if a discontinuity exists outside the LoS connecting the incident wave and the observer, it will not be directly accounted for in the cross-mapped temporal configuration. The reason is causality, where a discontinuity outside the LoS is not permitted to reflect waves back in time towards the observer. Special consideration must therefore be given to such cross-mapping configurations.

\begin{figure}[ht]
\centering
\subfloat[]{\includegraphics[width=2.25in]{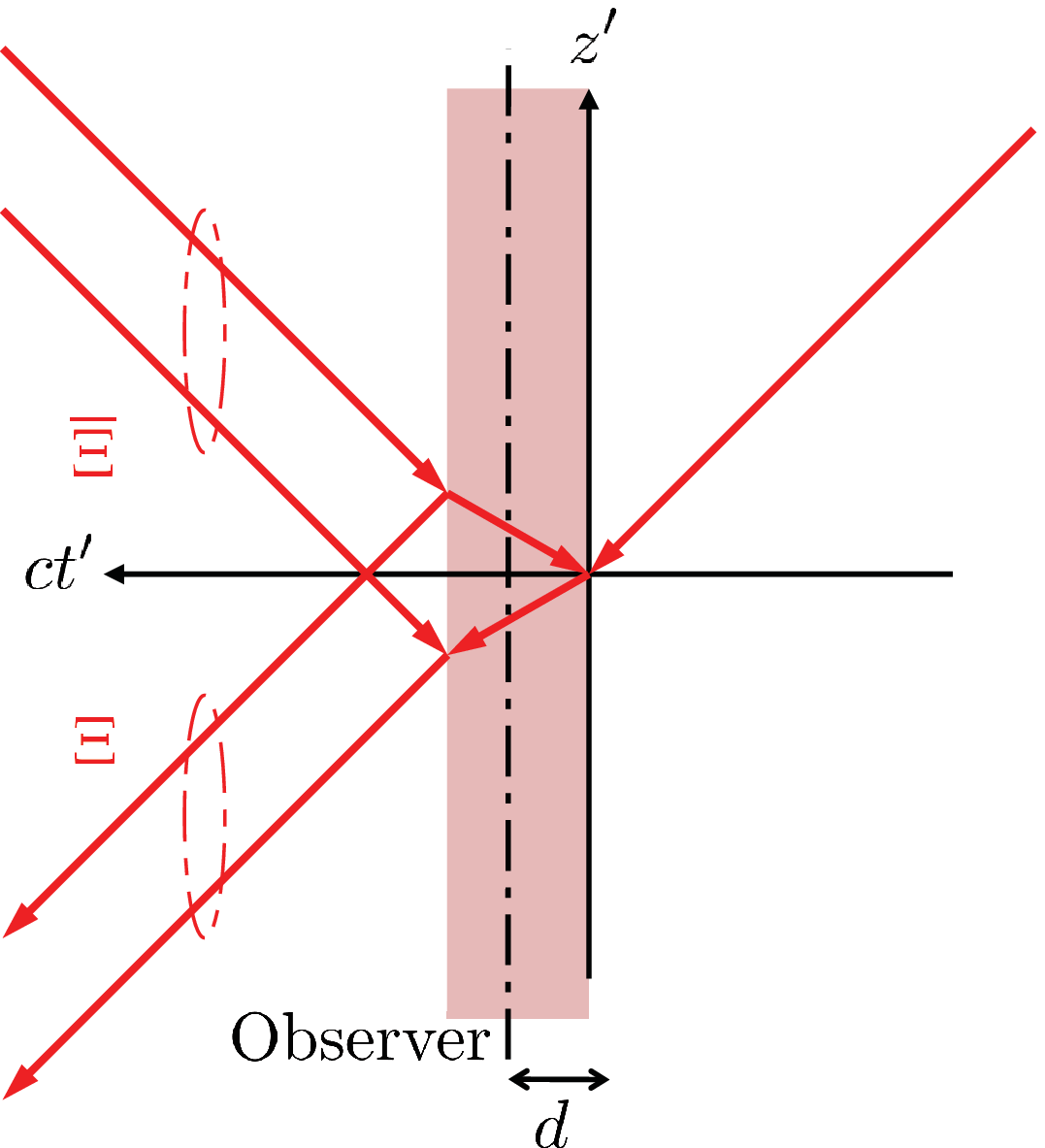}%
\label{fig:bcca}}
\\
\subfloat[]{\includegraphics[width=2.25in]{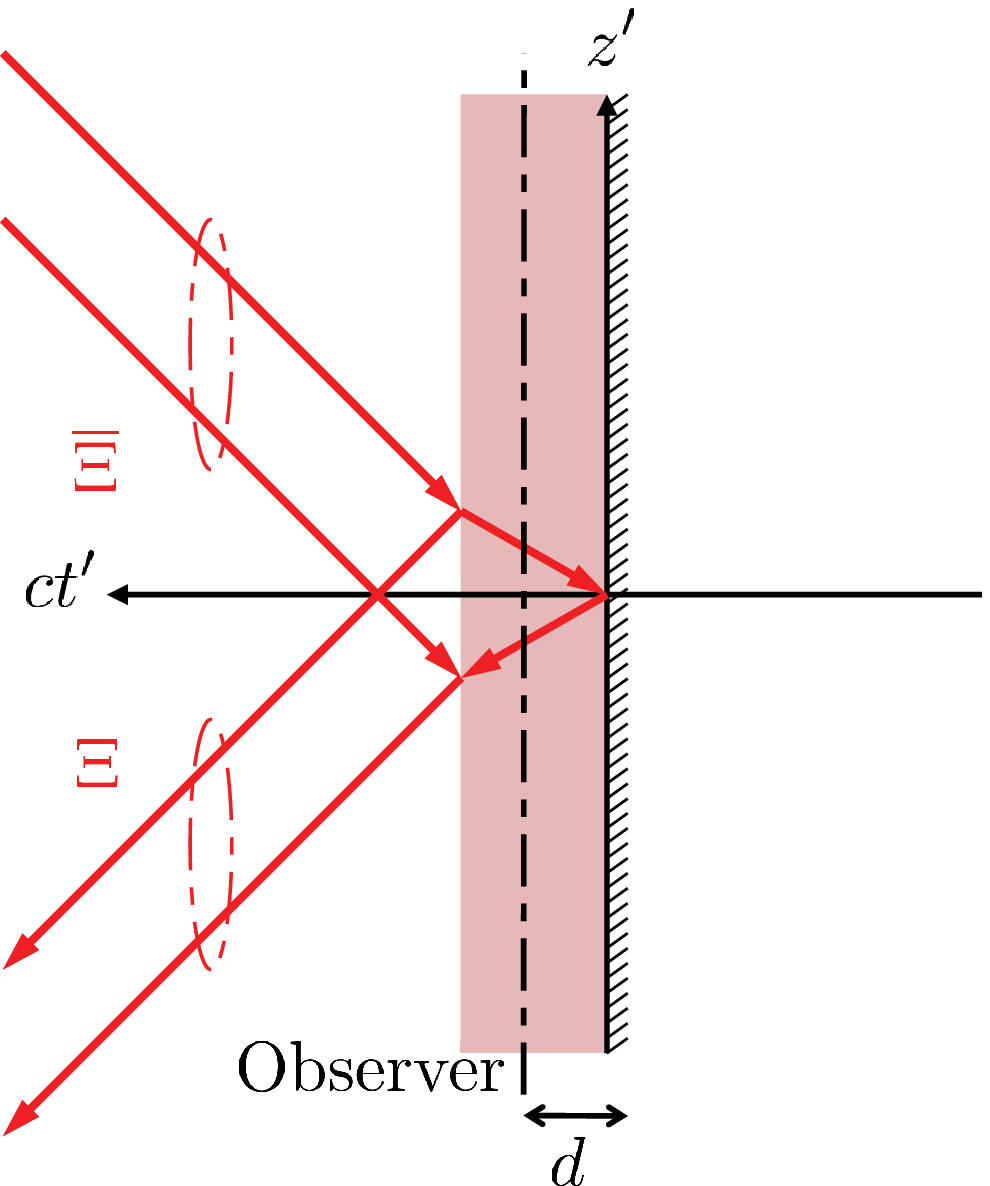}%
\label{fig:bccb}}
\caption{Spatial slab configuration with cross-mapped wave trajectories. \protect\subref{fig:bcca} Proper trajectory mapping starts from the radiation point at $z \rightarrow \infty$ if an open boundary exists. \protect\subref{fig:bccb} Proper trajectory mapping starts from a hard boundary behaving as a source.}
\label{fig:bcc}
\end{figure}

Consider the example of the dielectric slab of the previous section, where now the observer is located inside the slab at distance $d$ away from the exit interface, as shown in Fig.~\ref{fig:bcca}. Following the cross-mapping procedure outlined in the previous section and simply cross-mapping the trajectories from the incident wave to the observer yields here an incorrect result, since the second interface of the slab, formed in the future of the observer, is not accounted for. The proper cross-mapping has to be carried out between proper boundaries, namely interfaces where the values of the wave are prescribed. In this case, the boundaries are radiation boundaries located at $z \rightarrow \pm\infty$. This requirement is consistent with the nature of harmonic waves as they exist, by definition, in all space. The composite coefficients of the scattered wave at the observer are found by normalizing the scattered wave coefficients to $\Xi$ following Table~\ref{tab:crossmap} as%
\begin{align*}
\rho \left(d\right) &= e^{-i n_2 k d} \frac{\xi_{2,1}}{\Xi} = e^{-i n_2 k d} \frac{n_2 + n_1}{2 n_2} T,\\
\overline{\rho} \left(d\right) &= e^{i n_2 k d} \frac{\overline{\xi}_{2,1}}{\Xi} = e^{i n_2 k d} \frac{n_2 - n_1}{2 n_2} T,
\end{align*}
where the appropriate refractive index substitution has been performed, and the phase factors pertain to the location of the observer inside the slab.

Continuing with the observer inside the slab example, consider the case shown in Fig.~\ref{fig:bccb}, where the slab is backed by a hard boundary at its exit interface. The hard boundary is equivalent to a perfect electric conductor (PEC) boundary in the case of electromagnetic waves. Following the requirement that cross-mapping has to be carried out between boundaries, proper cross-mapping treats the hard boundary as a source that emits forward- and backward-propagating waves that exactly annihilate each other at the hard boundary. The cross-mapping procedure is otherwise carried out as in the previous case. The composite coefficients of the forward- and backward-propagating waves inside the slab are found as%
\begin{align*}
\rho \left(d\right) &= \frac{e^{-i n_2 k d}}{\xi_{1,2} e^{-i n_2 k l} + \overline{\xi}_{1,2} e^{i n_2 k l}}, \\
\overline{\rho} \left(d\right) &= \frac{e^{i n_2 k d}}{\xi_{1,2} e^{i n_2 k l} + \overline{\xi}_{1,2} e^{i n_2 k l}}.
\end{align*}

\subsection{Sources}

Waves emitted by a source always propagate away from the source, which in the one-dimensional configuration under consideration are the forward and backward directions. A discontinuity existing outside the LoS connecting the source and the observer will scatter the wave emitted by the source back towards the observer. In a cross-mapped temporal configuration, such scattered wave is not permitted to reflect back in time and hence will not reach the observer yielding incorrect evaluation of the waves at the observer.

Simple cross-mapping of waves emitted by a source is not possible due to causality, since the waves emitted by the source and propagate away from the observer, according to Sec.~\ref{sec:3a}, should be mapped to waves that propagate backward in time. Proper source modeling thus implies wave splitting at the source with an apriori undetermined source strength $\alpha$. The cross-mapping is carried out as previously described until the final boundary, whether it is a hard boundary or a radiation boundary. Applying the proper boundary condition at the final boundary determines the cross-mapped source strength $\alpha$. The corresponding source condition in the cross-mapped configuration is thus given by%
\begin{align}
\left[ G\left(z;z'\right) \right]_{z=z'^-}^{z=z'^+} &= 0, \label{eq:source_1}\\
\frac{\partial}{\partial z} \left[ G\left(z;z'\right) \right]_{z=z'^-}^{z=z'^+} &= \alpha, \label{eq:source_2}
\end{align}
where Eq.~(\ref{eq:source_1}) establishes the continuity of the Green's function, $G$, across the source, and Eq.~(\ref{eq:source_2}) establishes the jump of the derivative of $G$ across the source. In a spatial configuration, $\alpha = 1$, i.e. the source strength is always unity.

\begin{figure}[ht]
\centering
\subfloat[]{\includegraphics[width=2in]{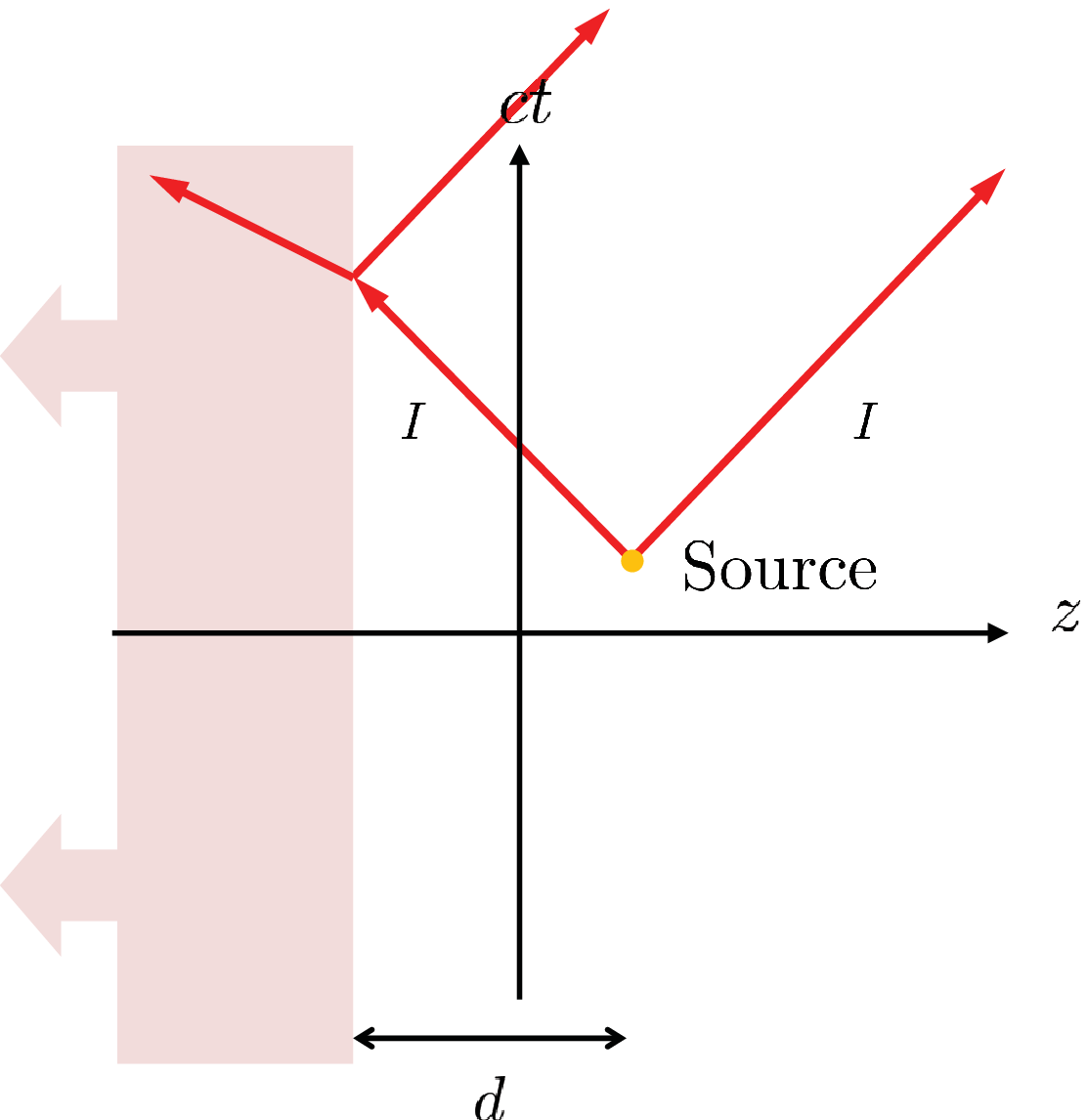}%
\label{fig:ssrc}}
\\
\subfloat[]{\includegraphics[width=2in]{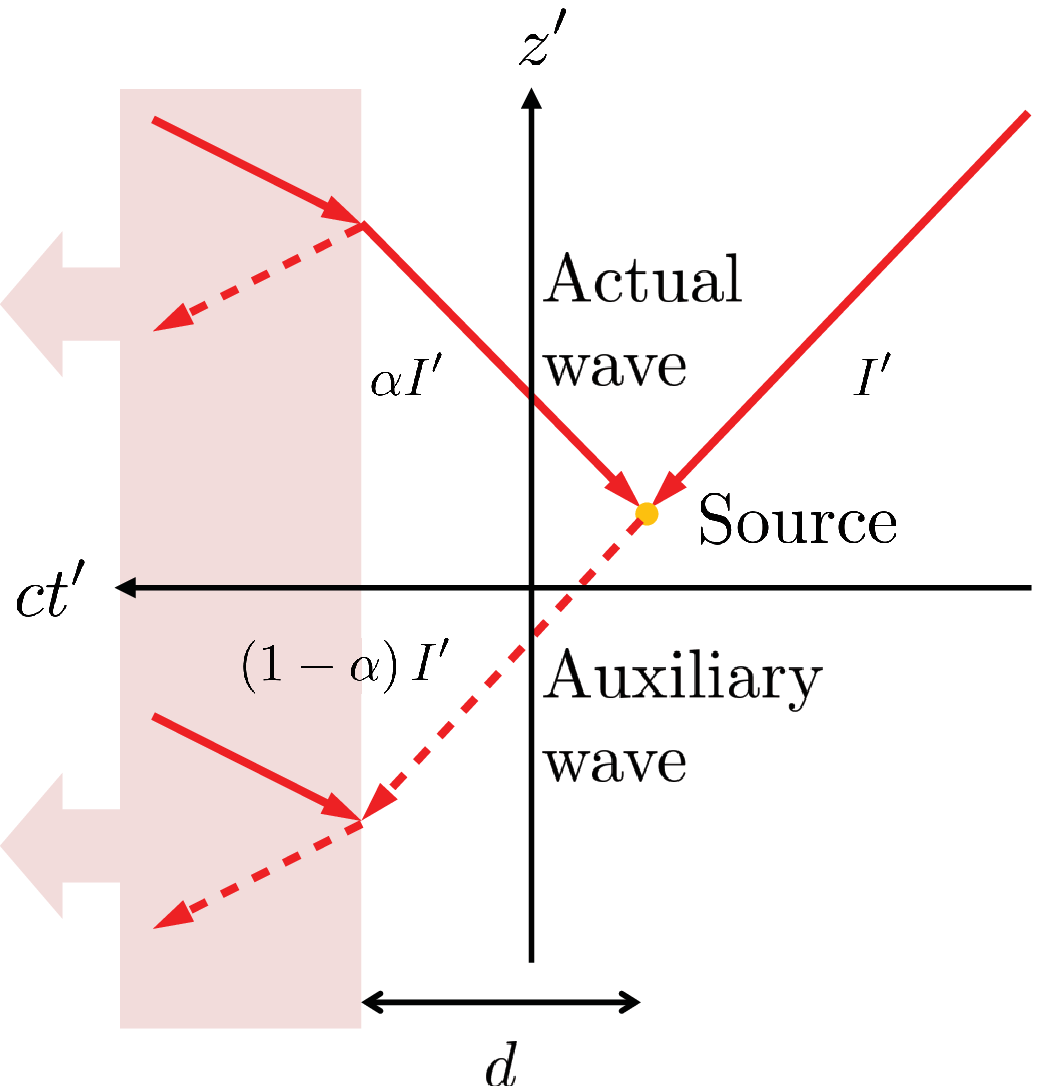}%
\label{fig:tsrc}}
\caption{A spatial configuration with a source located at a distance $d$ away from a half-space. \protect\subref{fig:ssrc} Wave trajectories in the spatial configuration, and \protect\subref{fig:tsrc} wave trajectories in cross-mapped configuration. In the cross-mapped configuration, the source splits in the incident wave into two waves, with an apriori undefined strength $\alpha$. The boundary condition at the end of the tracing process determines $\alpha$. The exit boundary condition in this configuration is the radiation condition in the dielectric at $ct' \rightarrow \infty$.}
\label{fig:src}
\end{figure}

An example involving a source in the cross-mapped configuration is shown in Fig.~\ref{fig:src}, where Fig.~\ref{fig:ssrc} is the spatial configuration consisting in a source located at a distance $d$ away from a dielectric half-space with refractive index $n$, and Fig.~\ref{fig:tsrc} is its corresponding cross-mapped configuration. In the cross-mapped configuration, the incident wave is split into two waves at the source, an actual wave and an auxiliary wave, with coefficients $\alpha$ and $1-\alpha$, respectively. The auxiliary wave is an anti-causal wave that is necessary for proper modeling, but has no physical significance otherwise. This auxiliary wave does not exist in the spatial configuration, hence its name. To determine $\alpha$, the radiation condition in the dielectric half-space implies the vanishing of all backward-propagating waves in the dielectric as%
\[
\alpha e^{i k d} \overline{\xi} + \left(1 - \alpha\right) e^{-i k d} \xi = 0,
\]%
which yields
\[
\alpha = \frac{1}{1 - \frac{\overline{\xi}}{\xi} e^{2 i k d}}.
\]%
The wave coefficient observed at $z = d+l$ away from the half-space is thus%
\[
T = \frac{e^{i k l}}{\alpha} = e^{i k l} \left[ 1 - \frac{\overline{\xi}}{\xi} e^{2 i k d} \right].
\]

Combining the proper source, continuity conditions and boundary conditions outside the LoS completes the proper formulation of the space-time cross-mapping method.

\section{Application in Transfer Matrix Method}
\label{sec:4}

The transfer matrix method (TMM) is a powerful computational method used to analyze the propagation of electromagnetic waves through a stratified medium~\cite{Berreman:72,Born:99}. However, when there are multiple interfaces, the multiple reflections between these interfaces may interfere destructively or constructively. The composite reflection or transmission through a multilayer structure is the sum of an infinite number of reflections, which scales up the computational burden.

\begin{figure}[ht]
\centering
\includegraphics[width=2.5in]{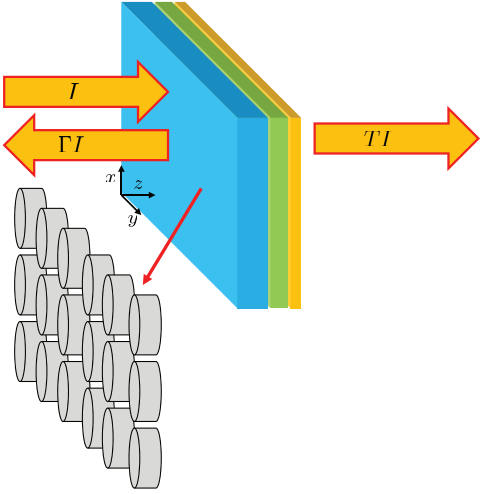}
\caption{Schematic of a layered medium or device. The incident field is designated by $I$, while the composite reflection and transmission coefficients are designated by $\Gamma$ and $T$, respectively. Notice that the cross-section of any given layer may be inhomogeneous as long as the layer is longitudinally invariant.}
\label{fig:tmm}
\end{figure}

The TMM takes advantage of the longitudinal invariance of each layer of the stratified medium with eigenmode expansion in the transverse cross-section, if necessary. Such expansion allows for analyzing inhomogeneous structures as long as they can be approximated by a stack of layers which may be inhomogeneous in the cross-section, while invariant in the longitudinal direction, as shown in Fig.~\ref{fig:tmm}. In such layers, the field at the end of the layer may be simply deduced from the field at the beginning of the layer through a simple phase accumulation. Additionally, the field continuity condition across the boundary between two layers may be expressed as a simple matrix operation. Scattering through a stack of layers may thus be represented as a matrix system, which is the product of the individual layer matrices~\cite{Lekner:87,Ko:88,Rumpf:11}. The computational burden increases with the increase in number of layers, hence employing a space-time cross-mapping may reduce this burden by increasing the computational speed or reducing the memory requirement.

\subsection{Transfer Matrix Method}

Application of the TMM requires each layer of medium to be longitudinally invariant. This invariance allows for the expansion of electromagnetic field in terms of plane-waves, the eigenmodes of the two-dimensional Cartesian plane, in the cross-section of each layer. This eigenmode expansion may be mathematically represented by $\partial_x \rightarrow i k_x$ and $\partial_y \rightarrow i k_y$, assuming the $z$-direction to be the longitudinal direction. Next, the transverse components of the wave vector are normalized as $\tilde{k}_\mathrm{j} = k_\mathrm{j}/k_0$, with $\mathrm{j} \in \{x,y\}$. The longitudinal distance is also normalized as $\tilde{z} = k_0 z$. Substituting these elements into Maxwell's equations and eliminating the longitudinal fields yields%
\begin{equation}
\label{eq:tmm_psi}
\frac{d}{d\tilde{z}}\mathbf{\Psi} - \mathbf{\Omega} \mathbf{\Psi} = 0,
\end{equation}
with the transverse fields vector%
\[
\mathbf{\Psi} = \left[\begin{array}{c}
E_x \\ E_y \\ H_x \\ H_y
\end{array}\right],
\]%
and the matrix%
\[
\mathbf{\Omega} = \left[\begin{array}{cccc}
0 & 0 & \frac{\tilde{k}_x \tilde{k}_y}{\epsilon_r} & \mu_r - \frac{\tilde{k}_x^2}{\epsilon_r} \\
0 & 0 &  \frac{\tilde{k}_x^2}{\epsilon_r} - \mu_r & -\frac{\tilde{k}_x \tilde{k}_y}{\epsilon_r} \\
\frac{\tilde{k}_x \tilde{k}_y}{\mu_r} & \epsilon_r - \frac{\tilde{k}_x^2}{\mu_r} & 0 & 0 \\
 \frac{\tilde{k}_x^2}{\mu_r} - \epsilon_r & -\frac{\tilde{k}_x \tilde{k}_y}{\mu_r} & 0 & 0
\end{array}\right], 
\]%
where $\epsilon_r$ and $\mu_r$ are respectively the effective relative permittivity and permeability of the layer considered.

Equation~(\ref{eq:tmm_psi}) is a matrix differential equation, whose solution takes the form%
\[
\mathbf{\Psi}\left(\tilde{z}\right) = \mathbf{W} e^{\lambda \tilde{z}} \mathbf{c},
\]
where $\mathbf{c} = \mathbf{W}^{-1}\mathbf{\Psi}(0)$, $\mathbf{\Psi}(0)$ is the field at the input interface of the layer, and $\mathbf{W}$ are there eigenvectors and $\lambda$ are eigenvalues of $\mathbf{\Omega}$.

The continuity conditions at the interface between two layers implies the continuity of the tangential fields as%
\[
\mathbf{\Psi}_{j+1}\left(0\right) = \mathbf{\Psi}_{j}\left(k_0 L_j\right),
\]
which may be expressed in terms of the eigenmodes of the layers as
\begin{equation}
\label{eq:tmm_cont}
\mathbf{c}_{j+1} = \mathbf{T}_j \cdot \mathbf{c}_{j},
\end{equation}
where $\mathbf{T}_j = \mathbf{W}_{j+1}^{-1} \mathbf{W}_j \exp(\lambda_j k_0 L_j)$ is the matrix that transfers the field from the layer $j$ into the layer $j+1$, and $L_j$ is the longitudinal length of the $j$ layer.

The system matrix is finally constructed from the transfer matrices of all layers as%
\begin{equation}
\label{eq:tmm_sys}
\mathbf{c}_T = \mathbf{T} \cdot \mathbf{c}_I,
\end{equation}
where $\mathbf{c}_I$ and $\mathbf{c}_T$ are the fields at the entry and the exit ports of the stratified medium, respectively, and $\mathbf{T} = \mathbf{T}_N \cdot \mathbf{T}_{N-1} \cdot \dots \mathbf{T}_1 \cdot \mathbf{W}_1^{-1} \mathbf{W}_I$ is the composite transfer matrix. Equation~(\ref{eq:tmm_sys}) describes an implicit scheme, since the unknown composite coefficients exist in $\mathbf{c}_I$ and $\mathbf{c}_T$. Solving the matrix equation~(\ref{eq:tmm_sys}) yields the composite reflection and transmission coefficients of the stratified medium.

\subsection{Space-Time Cross-Mapping in TMM}

Space-time cross-mapping may be advantageously integrated into the TMM with minimal modification and with the benefit of reducing the computation burden. The resulting method is a marching-on in space method, where the scattered field is deduced from the incident field without having to account for the infinite reflections between the layer interfaces.

To integrate the cross-mapping method into the TMM, the procedure outlined in Section~\ref{sec:3a} is followed and $\mathbf{c}_I$ and $\mathbf{c}_T$ are cross-mapped to $\tilde{\mathbf{c}}_T$ and $\tilde{\mathbf{c}}_I$, respectively. Next, the continuity condition~(\ref{eq:tmm_cont}) is replaced by the matrix generalization of Eqs.~(\ref{eq:xif_rec}) and~(\ref{eq:xib_rec}) as%
\begin{equation}
\label{eq:tmm_cm_cc}
\tilde{\mathbf{c}}_j = \tilde{\mathbf{T}}_j \cdot \tilde{\mathbf{c}}_{j+1},
\end{equation}%
where $\tilde{\mathbf{T}}_j = \tilde{\mathbf{W}}_{j}^{-1} \tilde{\mathbf{W}}_{j+1} \exp(\lambda_{j+1} k_0 L_{j+1})$ is the cross-mapped transfer matrix from layer $j+1$ to layer $j$, $\tilde{\mathbf{W}_j}$ are the eigenvectors corresponding to the cross-mapped configuration. The system matrix equation is constructed from Eq.~(\ref{eq:tmm_cm_cc}) as%
\begin{equation}
\label{eq:tmm_cm}
\tilde{\mathbf{c}}_T = \tilde{\mathbf{T}} \cdot \tilde{\mathbf{c}}_I,
\end{equation}
where $\tilde{\mathbf{T}}$ is the cross-mapped system transfer matrix. The marching-on in space nature of the cross-mapped configuration yields the implicit scheme in Eq.~(\ref{eq:tmm_cm}), where $\tilde{\mathbf{c}}_T$ is directly computed from $\tilde{\mathbf{c}}_I$, which contains all the composite scattered field coefficients. This is in contrast to Eq.~(\ref{eq:tmm_sys}), where the scattered field coefficients exist both in $\mathbf{c}_I$ and $\mathbf{c}_T$, and the composite scattered field coefficients are only available after solving the matrix system. Since the scattered field is obtained without the need to solve a linear matrix system in the cross-mapped TMM, the computational burden may be significantly reduced.

We should also note that the space-time cross-mapping method may be also applied to other techniques that analyze longitudinally invariant layers, such as rigorous coupled-wave analysis (RCWA)~\cite{Moharam:81} and plane-wave expansion method (PWEM)~\cite{Ho:90}. These techniques are inherently similar to the TMM, but they do not require homogenization of the layers of the medium.

\section{Conclusion}
\label{sec:5}

In this work, we have introduced the novel concept of space-time cross-mapping with application to wave scattering. The proposed cross-mapping allows for treating scattering in a space-varying configuration as scattering in a time-varying environment. Cross-mapping yields simplified expressions for the composite coefficients of the scattered wave by leveraging the asymmetry between space and time due to causality. Methods to properly account for boundaries outside the line-of-sight between the source and the observer, and to properly account for the presence of sources are presented. Space-time cross-mapping is not limited to harmonic waves and may be extended to pulsed wave through temporal spectrum composition. In addition, the presented cross-mapping for scalar waves may be directly extended to transverse electric-magnetic electromagnetic waves and longitudinally-invariant modes.

The proposed cross-mapping may have potential application to existing analytic and computational scattering schemes, such as the transfer matrix method, rigorous coupled wave analysis, and plane-wave expansion method. Cross-mapping may be also leveraged to decouple the scattering matrices in coupled filter design.

Future work may include dispersion analysis of the cross-mapped configuration for computational schemes, and extension of the method to three-dimensional configurations.


%

\end{document}